\documentclass{PoS}

\usepackage{fontenc}	
\usepackage{times}	
\usepackage{graphicx}	
\usepackage{amssymb}	

\usepackage{mathptmx}	

\newcommand{\ie}{$i.e.,\;$}
\newcommand{\eg}{$e.g.,\;$}

\title{Kpc-scale radio-jets in narrow-line Seyfert 1 galaxies}

\ShortTitle{Kpc-scale radio-jets in NLS1s}

\author{\speaker{Veeresh Singh}
\\
        Astronomy and Astrophysics Division, Physical Research Laboratory, Ahmedabad 380009, India\\
        E-mail: \email{veeresh@prl.res.in}}

\author{Hum Chand\\
       Aryabhatta Research Institute of Observational Sciences (ARIES), Manora Peak, Nainital 263002 India\\
       E-mail: \email{hum@aries.res.in}}
       
\author{C. H. Ishwara-Chandra\\
      National Centre for Radio Astrophysics, TIFR, Post Bag 3, Ganeshkhind, Pune 411007, India\\
       E-mail: \email{ishwar@ncra.tifr.res.in}}
       
\author{Preeti Kharb\\
       National Centre for Radio Astrophysics, TIFR, Post Bag 3, Ganeshkhind, Pune 411007, India\\
       E-mail: \email{kharb@ncra.tifr.res.in}}

\abstract{

Narrow-Line Seyfert 1 galaxies (NLS1s) are generally believed to be radio-quiet Active Galactic Nuclei (AGN) with relatively less-massive (10$^{5}$$-$10$^{8}$~M$_{\odot}$) 
Super-Massive Black Holes (SMBH). Using the FIRST radio detections of hitherto the largest sample of 11101 optically-selected NLS1s we find a rather uncommon 
population of 55 NLS1s with Kiloparsec-Scale Radio structures (KSRs). We note that the number of NLS1s with KSRs found in our study 
is only a lower limit considering the effects of sensitivity, spatial resolution and projection, and hence, the number of NLS1s with KSRs may be more common 
than thought earlier. The NLS1s with KSRs are distributed across a wide range of redshifts, flux densities and luminosities. NLS1s with KSRs tend to exhibit steep radio 
spectra and are likely to be misaligned version of blazar-like NLS1s. The ratio of IR to radio flux density 
(q$_{\rm 22~{\mu}m}$ $=$ log[S$_{\rm 22~{\mu}m}$/S$_{\rm 1.4~GHz}$]) versus radio luminosity plot suggests that 
KSRs are mostly powered by AGN, while KSRs in NLS1s with low radio luminosity (L$_{\rm 1.4~GHz}$ $<$ 10$^{23.5}$ W~Hz$^{-1}$) may have a contribution from circumnuclear 
starburst. The trend shown by KSRs in radio luminosity versus radio-size plot indicates that the radio-jets resulting in KSRs are either 
in the early phase of their evolution or inefficient to grow beyond the host galaxy.}

\FullConference{Revisiting narrow-line Seyfert 1 galaxies and their place in the Universe - NLS1 Padova\\
		9-13 April 2018 \\
		Padova Botanical Garden, Italy}

\begin{document}

\section{Introduction}

Seyfert 1 galaxies, a subclass of Active Galactic Nuclei (AGN), are generally characterized with the presence of broad emission lines 
(FWHM $\sim$ 5000 km s$^{-1}$) in their optical spectra. However, a fraction of Seyfert 1 galaxies termed 
as Narrow-Line Seyfert 1 galaxies (NLS1s) exhibit relatively 
narrower (FWHM $<$ 2000 km s$^{-1}$) emission lines. Unlike Broad-Line AGN (BL-AGN), NLS1s 
show strong permitted Fe II emission lines, relatively weaker forbidden emission lines 
({\eg}[O III]$\lambda$5007{\AA}/H${\beta}$ $<$ 3), steep soft X-ray spectra, and rapid X-ray variability \cite{Panessa11}. 
Furthermore, it is believed that, compared to BL-AGN, NLS1s host less-massive (10$^{5}$$-$10$^{8}$ M$_{\odot}$) Super-Massive Black Holes (SMBH) 
with relatively higher accretion rates \cite{Foschini15}.  
Radio observations show that, in general, NLS1s are radio-quiet {\ie}the ratio of radio flux density at 5 GHz to the 
monochromatic optical flux at 4400{\AA}  (R $=$ S$_{\rm 5~GHz}$/S$_{\rm 4400{\AA}}$) is less than 10 \cite{Komossa06}.
However, in recent years the detections of several NLS1s in $\gamma$-ray with Fermi/LAT have shown 
that at least a small fraction of NLS1s possess strong radio-jets similar to blazars \cite{Abdo09b,DAmmando15}.    
The existence of relativistic jets in many NLS1s is also confirmed by the direct imaging of parsec-scale radio jets using 
high-resolution Very Large Baseline Array (VLBA) observations \cite{Gu15}. 
We note that the high-resolution VLBA observations filter out the extended kiloparsec (kpc)-scale radio emission of low-surface-brightness. 
As per the orientation based unification model, the misaligned parent population of known blazar-like NLS1s with relativistic jets 
is expected to be several hundred. The relativistic radio-jets in misaligned sources are likely to 
give rise kpc-scale radio emission. However, in the literature, only a handful of NLS1s 
with KSRs are known \cite{Richards15,Berton18}. 
The scarcity of KSRs in NLS1s necessitates an investigation of kpc-scale radio emission in NLS1s using a statistically large sample.
In this paper we attempt to investigate the properties of KSRs in NLS1s. 
\section{Sample}
Our sample of 55 NLS1s with KSRs is extracted from \cite{Singh18}. 
Using 1.4 GHz Faint Images of the Radio Sky at Twenty-cm (FIRST), 1.4 GHz NRAO VLA Sky Survey (NVSS), 
327 MHz Westerbork Northern Sky Survey (WENSS), and 150 MHz TIFR GMRT Sky Survey (TGSS), 
\cite{Singh18} found the radio detections of only 498/11101 ($\sim$ 4.5$\%$) NLS1s in the sample of 
11101 optically-selected NLS1s reported in \cite{Rakshit17}.  
The identification method of KSRs in FIRST radio sources includes - (i) segregation of resolved and unresolved radio sources 
by accounting for the errors involved in fitting an elliptical Gaussian to the single component radio source, and then (ii) among the resolved 
sources a further criterion of ${\rm S_{int}/S_{peak}}$ $>$ 1.12 is applied to identify KSRs. 
A detailed description on the identification of KSRs and unresolved sources can be found in \cite{Singh18}.  
%
%
%
\section{Radio properties of NLS1s with KSRs}
Table~\ref{table:HistProp} lists the range, median and standard deviation values of various distributions ({\ie}projected linear radio size, redshift, 
1.4 GHz flux density, 1.4 GHz radio luminosity, radio spectral index, radio-loudness parameter and 
q$_{\rm 22~{\mu}m}$) for the NLS1s with KSRs and the NLS1s with no detected KSRs.   
\begin{table*}
\begin{minipage}{140mm}
\caption{Radio properties of our sample of NLS1s.}
\begin{tabular}{@{}lccccr@{}}
\hline
Parameter             & Unit                & \multicolumn{2}{c}{NLS1s with KSRs}           &  \multicolumn{2}{c}{NLS1s with no detected KSRs}        \\    
                      &                     &   Range (median, std)           &     No.      &    Range (median, std)        &     No.       \\ \hline
Radio-size             & (kpc)              & 1.13$-$1113.0 (11.9, 203)       &      55      &        ...               &      443           \\  
Redshift              &                     & 0.0098$-$0.7977 (0.26, 0.22)    &     55       & 0.031$-$0.7969 (0.36, 0.22)  &    443         \\
S$_{\rm 1.4~GHz}$     &      (mJy)          & 1.71$-$3733.7 (5.56, 496)       &     55       & 1.0$-$8360.6 (2.31, 402.1)   &    443         \\
logL$_{\rm 1.4~GHz}$  & (W Hz$^{-1}$)       & 21.85$-$26.48 (24.05, 1.34)     &      55      & 21.89$-$27.14 (24.0, 1.01)   &    443         \\
${\alpha}_{\rm 150~MHz}^{\rm 1.4~GHz}$&     &  -1.19$-$-0.31 (-0.62, 0.22)    &      18      &  -1.13$-$0.57 (-0.49, 0.35)  &   61           \\
logR$_{\rm 1.4~GHz}$   &                    &  0.42$-$4.67 (1.64, 1.06)       &      55      & -0.02$-$5.39 (1.46, 0.88) &   443          \\ 
q$_{\rm 22~{\mu}m}$    &                    &  -2.39$-$1.61 (1.04, 0.90)      &      38      & -2.40$-$2.10 (1.04, 0.63) &   316           \\
\hline  
\end{tabular}
\label{table:HistProp} 
\end{minipage}
\\
{\it Notes} - std : standard deviation of the distribution. Spectral index between 150~MHz and 1.4~GHz is derived by assuming 
a power law (S$_{\nu}$ $\propto$ ${\nu}^{\alpha}$) radio spectrum.
\end{table*}
\begin{figure*}
\centering
\includegraphics[angle=0,width=6.5cm,trim={0.75cm 0.75cm 0.75cm 0.75cm},clip]{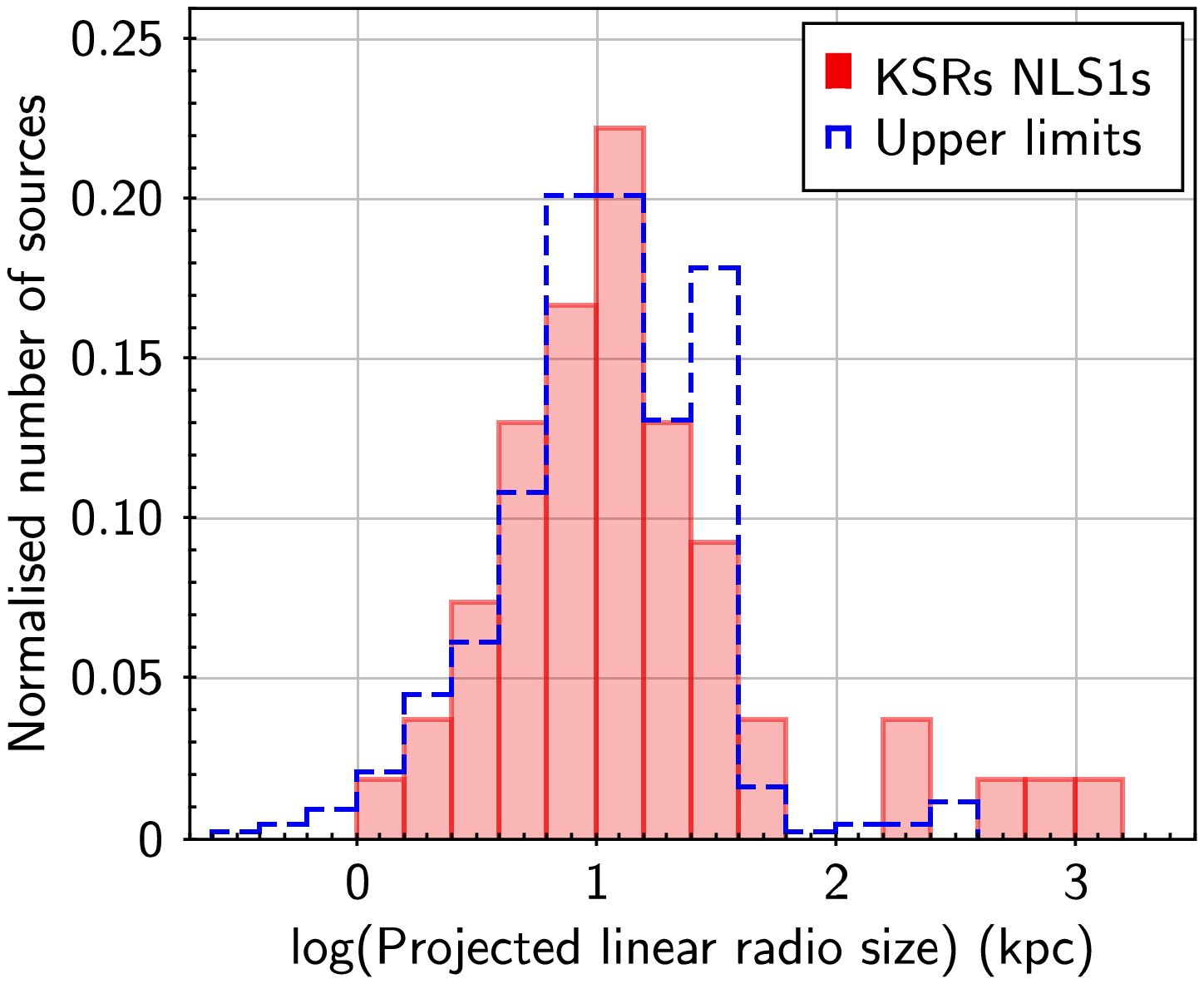}{\includegraphics[angle=0,width=6.5cm,trim={0.75cm 0.75cm 0.75cm 0.75cm},clip]{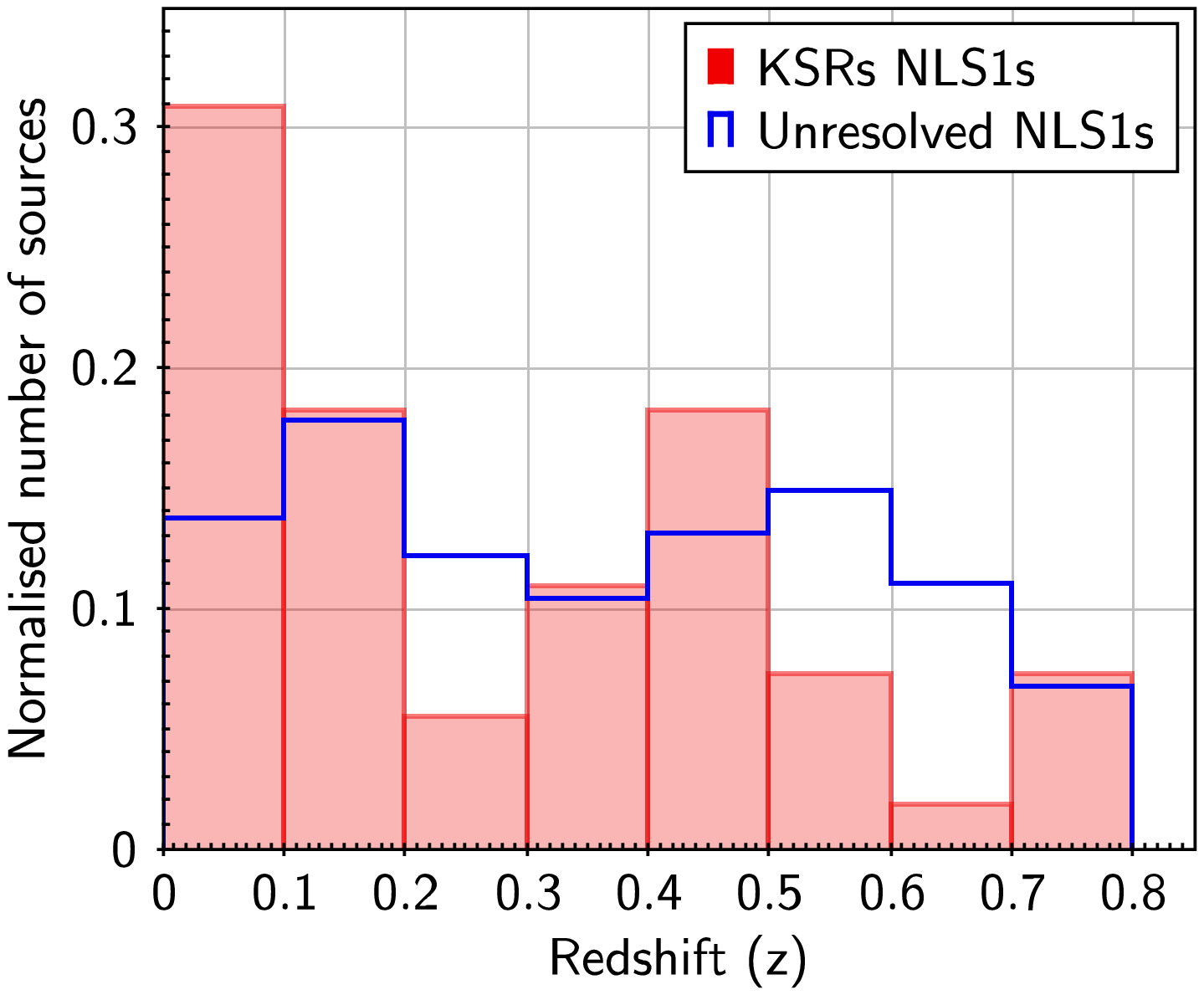}}
\caption{{\it Left panel} : The histogram of projected linear radio-size of KSRs and upper limits for unresolved sources. 
{\it Right panel} : Redshift distributions of KSRs and unresolved sources.} 
\label{fig:SizeRedshiftHist} 
\end{figure*}
\begin{figure*}
\centering
\includegraphics[angle=0,width=6.5cm,trim={0.75cm 0.75cm 0.75cm 0.75cm},clip]{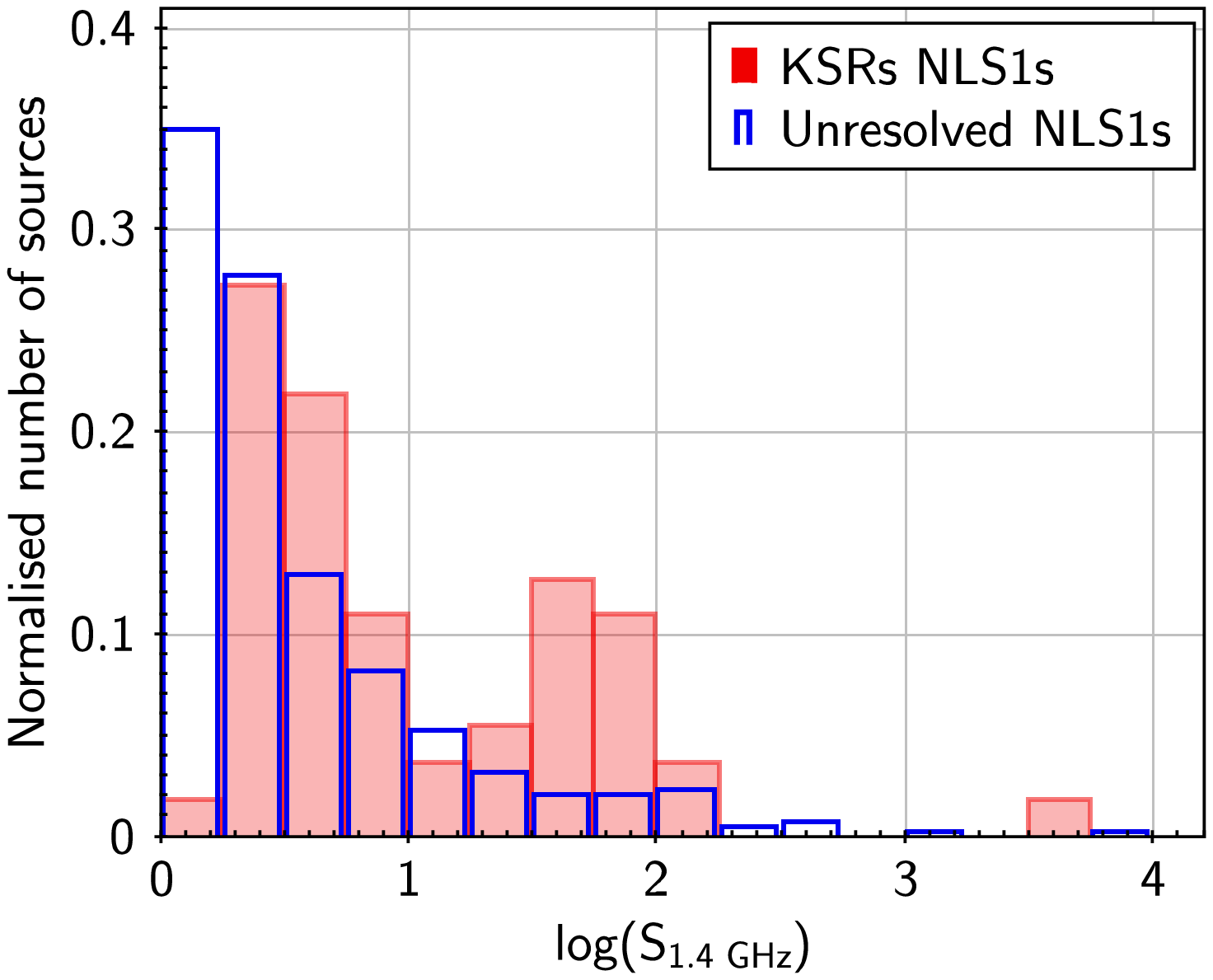}{\includegraphics[angle=0,width=6.5cm,trim={0.75cm 0.75cm 0.75cm 0.75cm},clip]{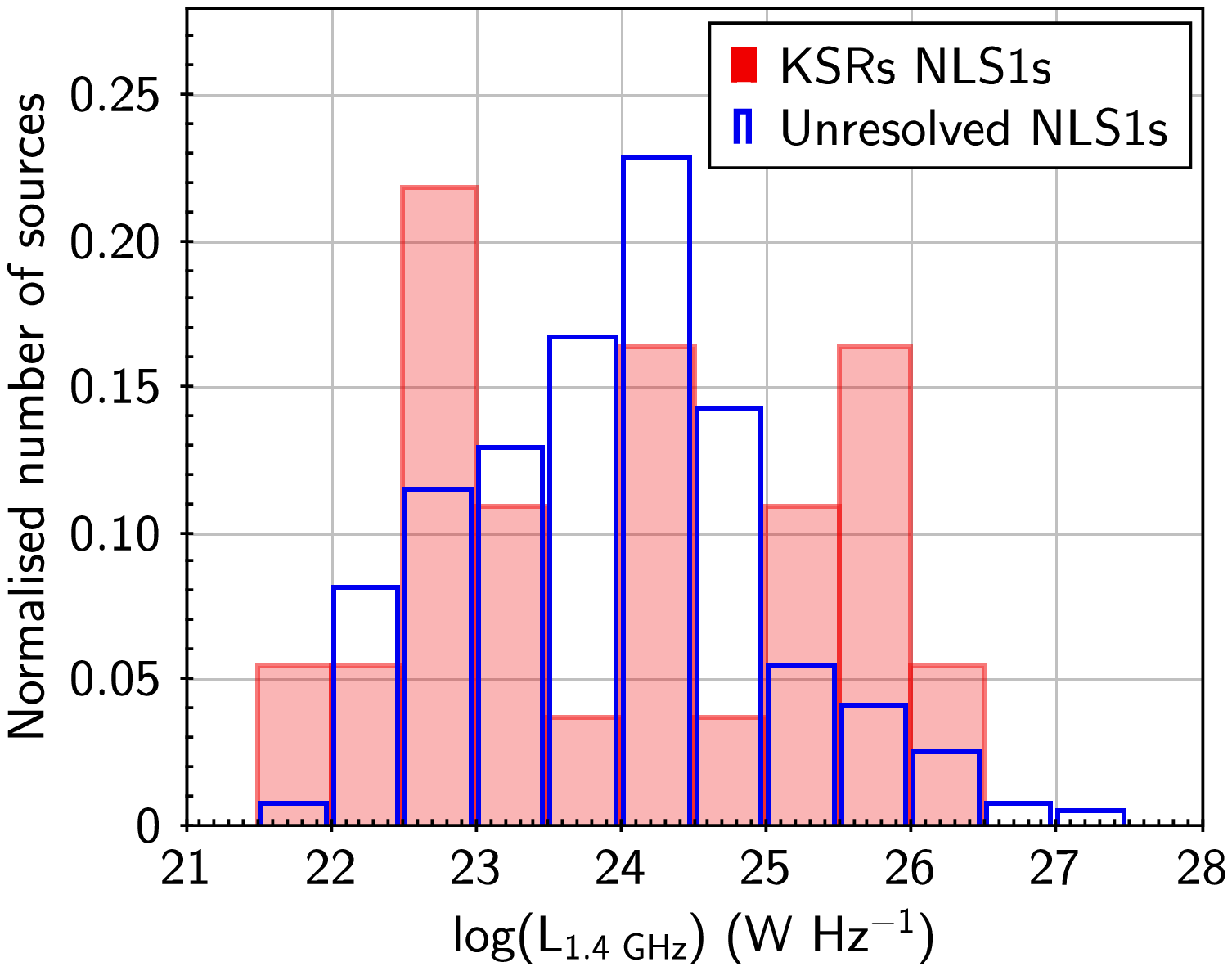}}
\caption{{\it Left panel} : 1.4 GHz flux density distributions. {\it Right panel} : 1.4 GHz radio luminosity distributions of KSRs and unresolved sources.} 
\label{fig:FluxLuminHist} 
\end{figure*}
%
\begin{figure*}
\centering
\includegraphics[angle=0,width=6.5cm,trim={0.75cm 0.75cm 0.75cm 0.75cm},clip]{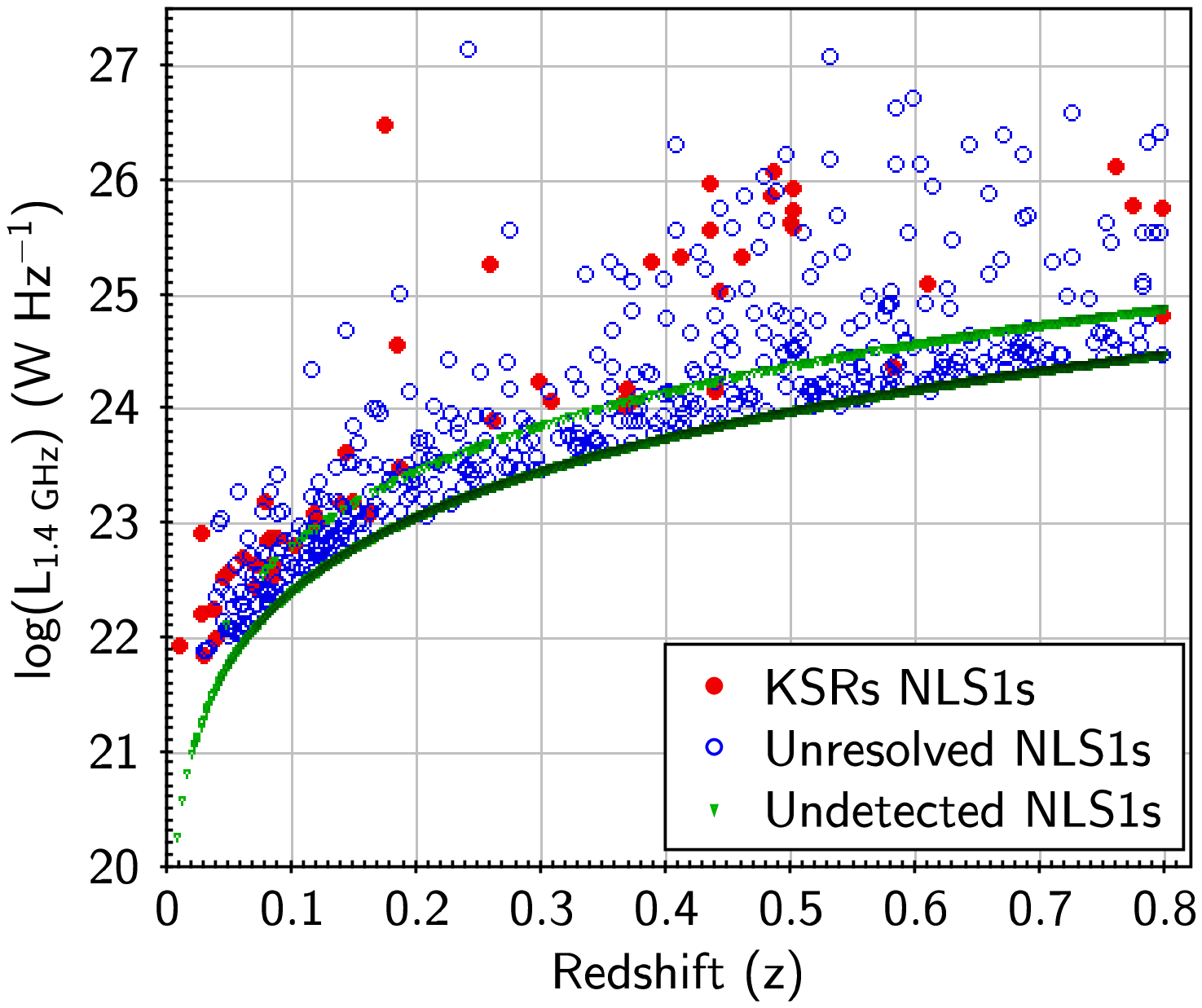}{\includegraphics[angle=0,width=6.5cm,trim={0.75cm 0.75cm 0.75cm 0.75cm},clip]{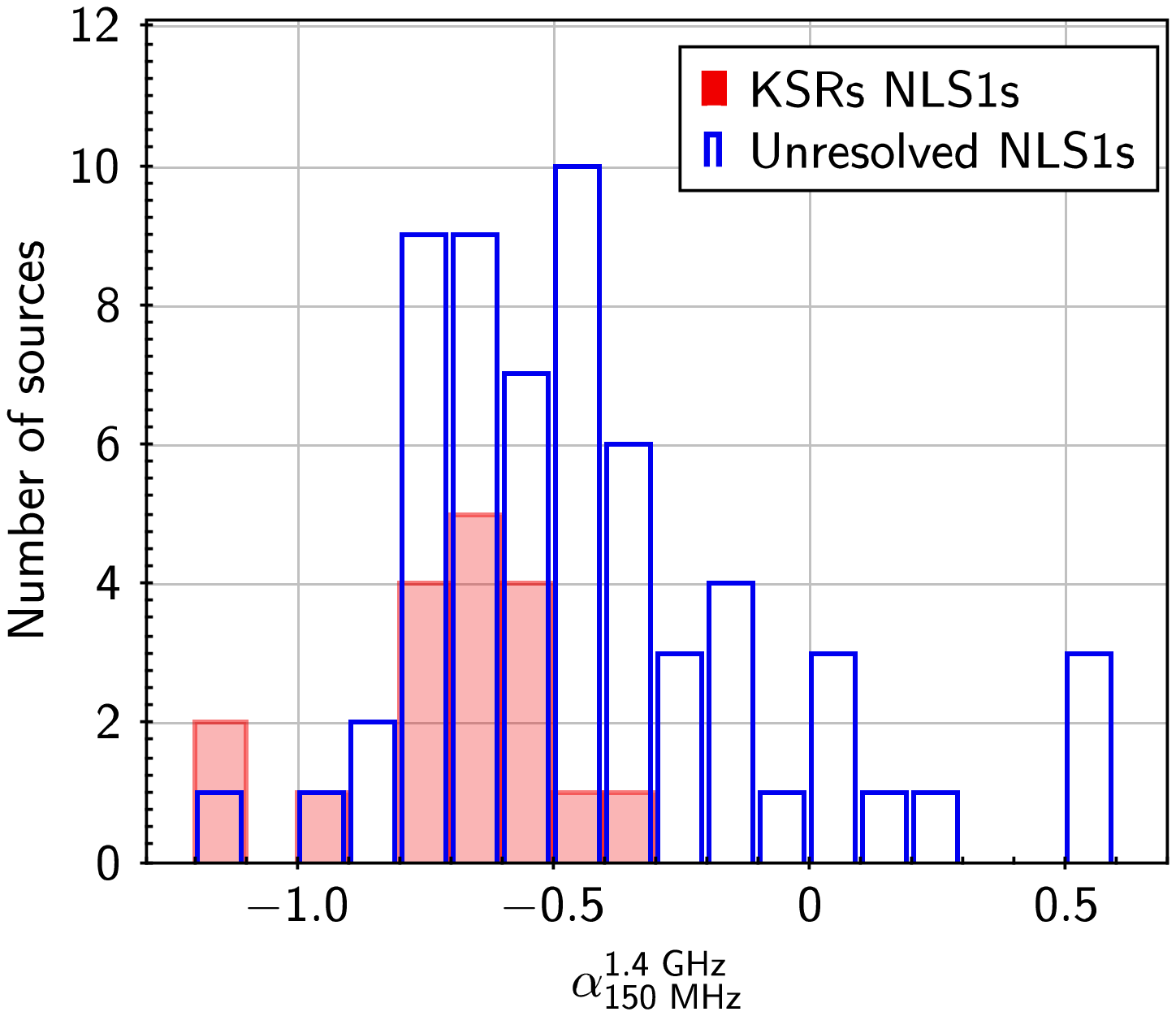}}
\caption{{\it Left panel} : Redshift versus 1.4 GHz luminosity. The upper and lower green curves represent upper limits from NVSS and FIRST survey, respectively. 
{\it Right panel} :  Two-point (150~MHz$-$1.4~GHz) spectral index distributions of KSRs and unresolved sources.}
\label{fig:RedshiftVsLuminSpIn} 
\end{figure*}
%
%
%
\begin{figure*}
\centering
\includegraphics[angle=0,width=5.0cm,trim={0.75cm 0.75cm 1.0cm 1.0cm},clip]{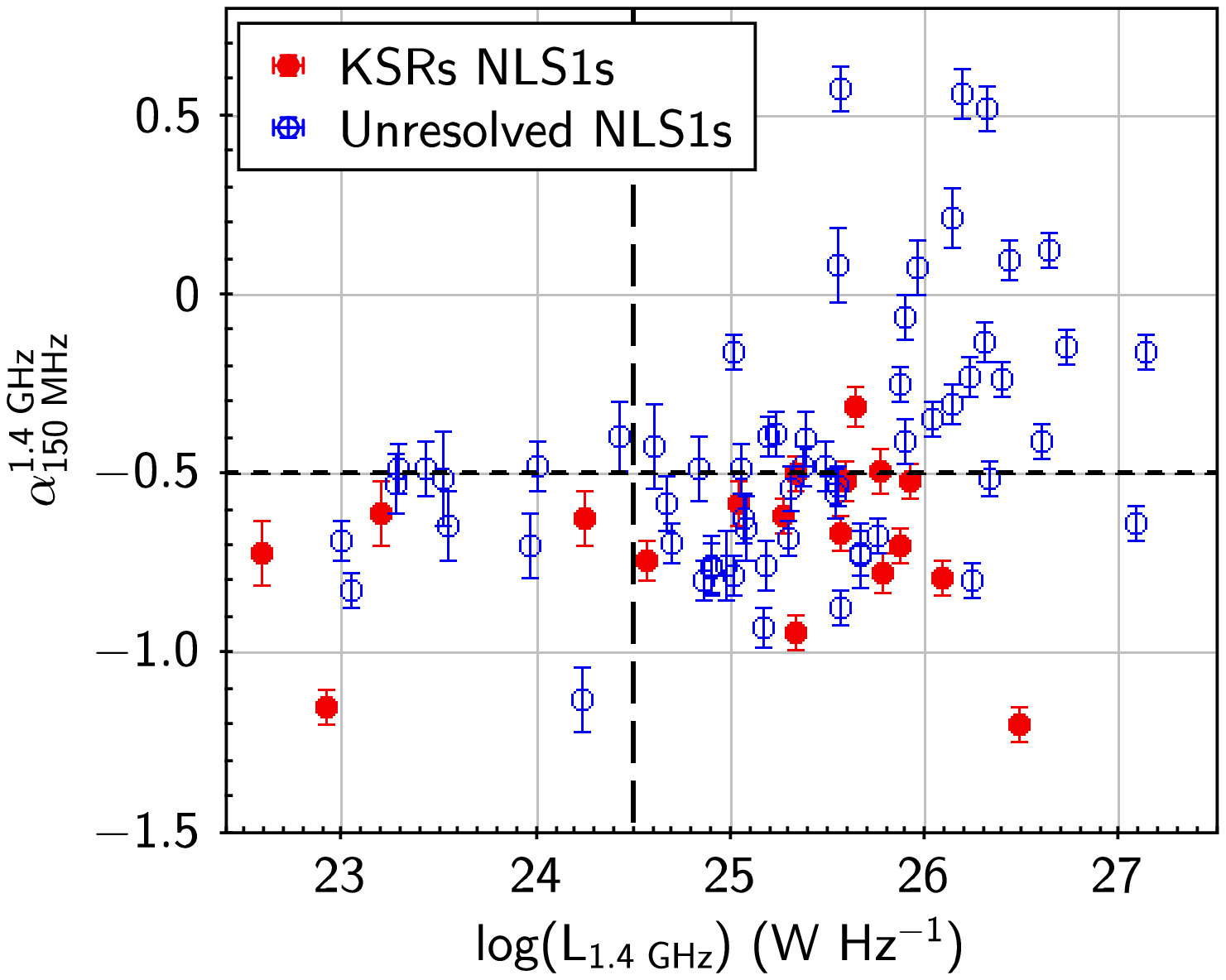}{\includegraphics[angle=0,width=5.0cm,trim={0.75cm 0.75cm 1.0cm 1.0cm},clip]{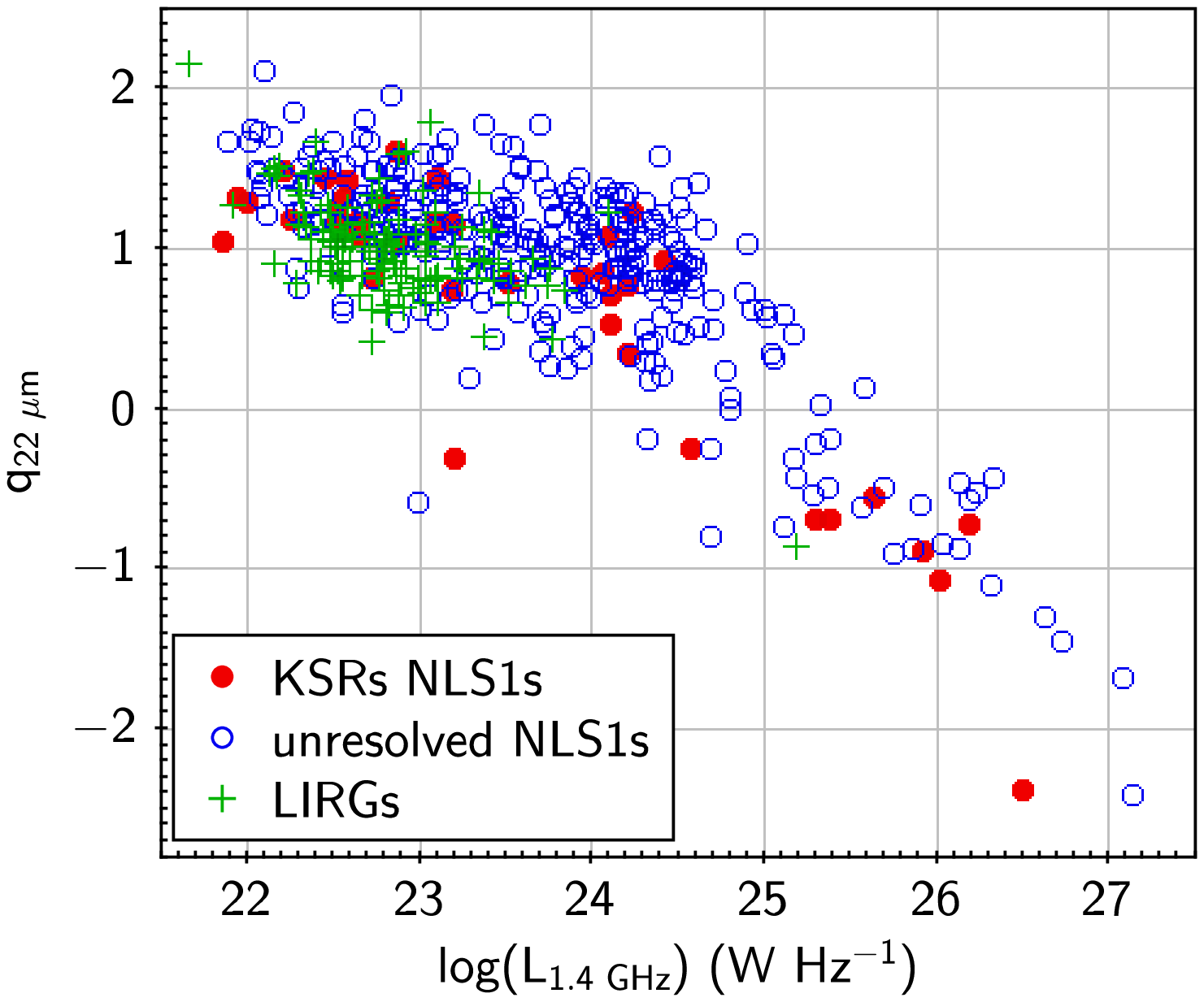}}
{\includegraphics[angle=0,width=5.0cm,trim={0.75cm 0.75cm 1.0cm 1.0cm},clip]{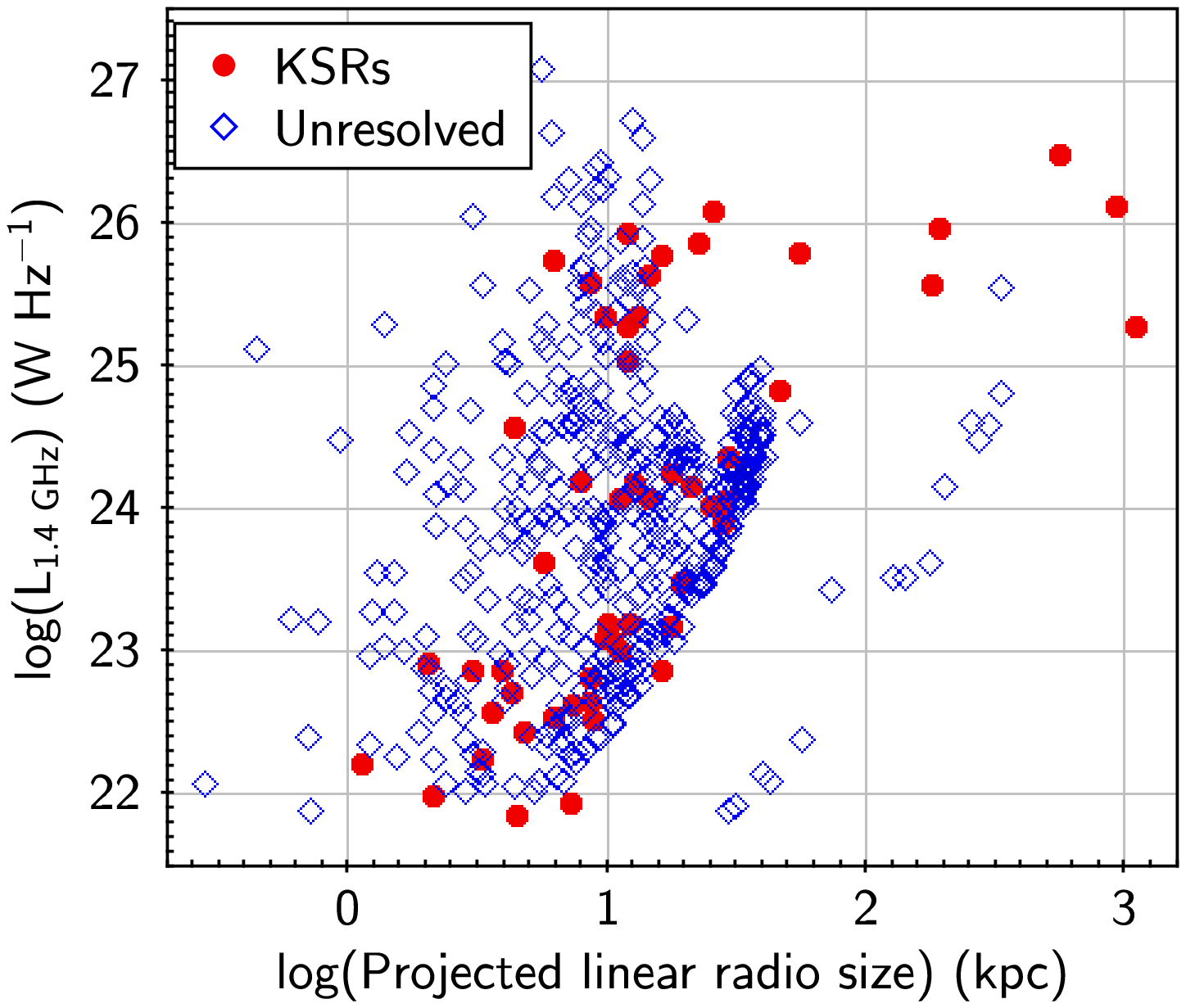}}
\caption{{\it Left panel} : 1.4 GHz radio luminosity versus two-point spectral index. {\it Middle panel} : q$_{\rm 22{\mu}m}$ versus 1.4 GHz radio luminosity. 
{\it Right Panel} : Projected linear radio-size versus 1.4 GHz radio luminosity plot.} 
\label{fig:SpInVsL} 
\end{figure*}
%
\subsection{Kiloparsec-Scale Radio structures (KSRs) in NLS1s}
We estimate the radio-sizes of 55 NLS1s with resolved radio emission and upper limits for 443 NLS1s with unresolved radio emission using FIRST observations whenever available, 
otherwise NVSS measurements are used. 
We find that 55 NLS1s with resolved radio emission have radio-size ranging from 1.1 kpc to 1.1 Mpc with a median of 11.3 kpc, hence confirming 
that a small fraction (55/11101 $\sim$ 0.5$\%$) of NLS1s possess KSRs (see Fig.~\ref{fig:SizeRedshiftHist}, left panel and Table~\ref{table:HistProp}). 
We note that the number of NLS1s with KSRs found in our study is only a lower limit as sources appearing unresolved, particularly at higher redshifts, 
may possess KSRs and jet-like KSRs aligned close to the line-of-sight would appear foreshortened due to projection effect.    
Notably, our study reveals the highest number of NLS1s with KSRs till date. 
It is worth noting that barring a few NLS1s with radio-size $\geq$ 100 kpc, most of the NLS1s have radio-sizes (ranging from 1.1 kpc to 30 kpc) smaller 
than the average optical size of the host galaxy. Thus, the radio-size of KSRs in NLS1s seem similar to those found in Seyfert galaxies \cite{Singh15a}.   
Previous studies have shown that the radio emission in NLS1s is mostly parsec-scale compact \cite{Gu15} and KSRs are rare \cite{Richards15}. 
Although, recently using deep (5$\sigma$ $\sim$ 50 $\mu$Jy) and high-resolution (0$^{{\prime}{\prime}}$.5) 5~GHz VLA A-configuration radio observations \cite{Berton18} 
have found KSRs in 21 sources among the sample of 74 NLS1s. 
Therefore, deep arcsec-resolution radio observations of our NLS1s can detect more KSRs.     
%
%
%
\subsection{Redshift and 1.4 GHz radio flux densities}
The redshift distribution 55 NLS1s with KSRs spans over 0.0098 to 0.7977 with a median of 0.26, 
while 443 NLS1s with unresolved radio counterparts are distributed over the same range but with 
a higher median of 0.36 (see Fig.~\ref{fig:SizeRedshiftHist}, right panel). 
It is evident that KSRs are preferentially found in NLS1s at lower redshifts, which 
can be due to an observational bias {\ie}the radio structures of similar scale tend to have smaller 
apparent angular sizes with the increase in redshift. 
%
%
We note that the KSRs are found in both faint as well as bright radio sources. 
The 1.4~GHz flux densities of NLS1s with KSRs range from 1.71 mJy to 3733.7 mJy with a median of 5.56 mJy, while 1.4~GHz flux densities of NLS1s with unresolved 
emission span over a much larger range of 1.0 mJy to 8360.6 mJy but with a lower median of 2.31 mJy (see Fig.~\ref{fig:FluxLuminHist}, left panel). 
Thus, most of the sources at the very faint end (S$_{\rm 1.4~GHz}$ $\leq$ 2.5 mJy) appear unresolved. 
%
\subsection{Radio luminosities and spectra}
1.4 GHz radio luminosity distribution for both unresolved and KSRs NLS1s span over a wide range (10$^{21}$ $-$ 10$^{27}$ W Hz$^{-1}$) with a median 
of $\sim$ 10$^{24}$ W Hz$^{-1}$ (see Fig.~\ref{fig:FluxLuminHist}, right panel). Moreover, unlike unresolved radio sources, the radio luminosity distribution 
of KSRs sources shows three apparent peaks at low (L$_{\rm 1.4~GHz}$ = 10$^{22.75}$ W~Hz$^{-1}$), moderate ((L$_{\rm 1.4~GHz}$ = 10$^{24.25}$ W~Hz$^{-1}$) 
and high ((L$_{\rm 1.4~GHz}$ = 10$^{25.75}$ W~Hz$^{-1}$) luminosity. We note that, on average, KSRs sources with low, moderate and high radio luminosities are found at 
low ($z_{\rm median}$ $\sim$ 0.1), moderate ($z_{\rm median}$ $\sim$ 0.3) and high redshift ($z_{\rm median}$ $\sim$ 0.5), 
respectively (see Fig.~\ref{fig:RedshiftVsLuminSpIn}, left panel). The prevalence of radio powerful sources at higher redshifts is consistent with the fact 
that co-moving density of radio-powerful AGN increases with redshift over $z$ $\sim$ 0.0$-$2.0.  
However, three peaks in the radio luminosity distribution of KSR sources appear intriguing and deep radio observations of a large 
sample of NLS1s are required to confirm the existence of these peaks. 
Further, we note that the KSRs are present in both radio-quiet and radio-loud NLS1s. 
Although, the median value of radio-loudness for NLS1s with KSRs is marginally higher than that for the NLS1s with unresolved radio emission (see Table~\ref{table:HistProp}).
\\
We estimate two-point spectral index (${\alpha}_{\rm 150~MHz}^{\rm 1.4~GHz}$) for KSRs and unresolved sources using 1.4 GHz flux density from FIRST or NVSS 
and 150 MHz flux density from TGSS. 
The distribution of ${\alpha}_{\rm 150~MHz}^{\rm 1.4~GHz}$ shows that spectral indices for 18 KSR sources range from -1.19 to -0.31 with a median of -0.61, 
while 61 unresolved radio sources have spectral indices spanning over -1.13 to 0.57 with a median of -0.49 (see Fig.~\ref{fig:RedshiftVsLuminSpIn}, right panel). 
The steeper radio spectra of KSRs are consistent with the optically-thin synchrotron radio emission, while 
compact sources can have flat/inverted radio spectrum due to free-free or synchrotron self-absorption. 
The very steep (${\alpha}_{\rm 150~MHz}^{\rm 1.4~GHz}$ $<$ -1.0) radio spectra can also be indicative of the relic radio emission caused by the previous 
episode of AGN activity \cite{Kharb16,Congiu17}. 
From luminosity versus spectral index plot (Fig.~\ref{fig:SpInVsL}, left panel) it is apparent that the KSRs sources distributed across all luminosities 
have steep spectra, while unresolved sources show a clear trend of flattening of radio spectra at higher luminosities. 
The unresolved sources with high radio luminosity and flat/inverted radio spectra can be blazar-like sources, in which 
luminosity is boosted due to relativistic beaming effect.  
\subsection{Origin and evolution of KSRs in NLS1s}
We attempt to investigate whether KSRs are powered by AGN or star-formation by using the radio-IR correlation which can also be represented as 
the ratio of IR to radio flux density {\ie}q$_{\rm 22~{\mu}m}$ = log[S$_{\rm 22~{\mu}m}$/S$_{\rm 1.4~GHz}$], where IR flux at 22 $\mu$m is 
taken from Wide-field Infrared Survey Explorer (WISE). 
Star-forming galaxies are known to show a tight correlation between IR and radio emission while radio-loud AGN tend to deviate. 
In Fig.~\ref{fig:SpInVsL} (right panel) we plot q$_{\rm 22~{\mu}m}$ versus 1.4 GHz radio luminosity for NLS1s with KSRs and with unresolved radio 
emission and Luminous InfraRed Galaxies (LIRGs). It is evident that q$_{\rm 22~{\mu}m}$ decreases with the increase in radio luminosity. 
All the radio-luminous NLS1s (L$_{\rm 1.4~GHz}$ $>$ 10$^{23.5}$ W~Hz$^{-1}$) have q$_{\rm 22~{\mu}m}$ much lower than that for star-forming LIRGs. 
Therefore, KSRs in radio-luminous NLS1s is likely to be driven by AGN. However, a fraction of NLS1s with lower radio luminosities show q$_{\rm 22~{\mu}m}$ similar 
to LIRGs, and hence, KSRs in these NLS1s can have a contribution from star-formation. 
Although, the small radio sizes of NLS1s with low radio luminosity (Fig.~\ref{fig:SpInVsL}, right panel) indicate that the radio emission is driven 
by either AGN or a compact circumnuclear starburst. 
\\
From Fig.~\ref{fig:SpInVsL} (right panel) it is apparent that radio-size increases with the increase in radio luminosity except for very large 
evolved radio sources ($>$ 100 kpc). The trend shown by KSRs in radio luminosity versus projected linear radio-size is consistent 
with the dynamical evolutionary models, suggesting that the compact parsec-scale radio-jets evolve into kpc-scale radio-jets which
eventually turn into large radio galaxies displaying hundreds of kpc jet-lobe structures \cite{Snellen2000}. 
Therefore, the radio jets of a few kpc scale in NLS1s may be in the early phase of evolution similar to Giga-hertz Peaked Spectrum (GPS) 
and Compact Steep Spectrum (CSS) radio sources \cite{Caccianiga17,Berton16}. 
However, the lack of large KSRs ($>$ 100) in NLS1s infers that radio-jets in many NLS1s may not be sufficiently powerful to dispel the 
Inter-Stellar Medium (ISM), and hence, may remain confined within the host galaxy.    
\section{Conclusions} 
Using FIRST radio detections of hitherto the largest optically-selected sample of 11101 NLS1s we obtain the largest sample of 55 NLS1s with KSRs. 
The NLS1s with KSRs are distributed across a wide range of redshifts, flux densities and luminosities. 
Unlike unresolved radio sources, the KSRs tend to show steep radio spectra, therefore NLS1s with KSRs are likely to be misaligned version of the 
flat-spectrum radio-loud NLS1s. The KSRs are mostly powered by AGN while they can have a contribution from circumnuclear starburst in NLS1s with 
low radio luminosities. The radio luminosity versus radio size plot suggests that the radio-jet driving KSRs in NLS1s are either 
in the early phase of their evolution or are frustrated by the ISM.   
\section*{Acknowledgements}
VS acknowledges support from PRL Ahmedabad, and the RadioNet. 
This conference has been organized with the support of the Department of Physics and Astronomy ``Galileo Galilei'', the 
University of Padova, the National Institute of Astrophysics INAF, the Padova Planetarium, and the RadioNet consortium. 
RadioNet has received funding from the European Union's Horizon 2020 research and innovation programme under 
grant agreement No~730562.

%

\bibliographystyle{JHEP}
\bibliography{NLSY1}

\end{document}